\DocumentMetadata{}
\documentclass[sigconf]{acmart}
\usepackage{multirow}
\usepackage{makecell}
\usepackage{graphicx}
\usepackage{textcomp}
\AtBeginDocument{%
  \providecommand\BibTeX{{%
    \normalfont B\kern-0.5em{\scshape i\kern-0.25em b}\kern-0.8em\TeX}}}

\copyrightyear{2026}
\acmYear{2026}
\setcopyright{cc}
\setcctype{by}
\acmConference[ICMR '26]{International Conference on Multimedia Retrieval}{June 16--19, 2026}{Amsterdam, Netherlands}
\acmBooktitle{International Conference on Multimedia Retrieval (ICMR '26), June 16--19, 2026, Amsterdam, Netherlands}
\acmDOI{10.1145/3805622.3810672}
\acmISBN{979-8-4007-2617-0/2026/06}

\begin{document}

\title{A Decomposed Retrieval-Edit-Rerank Framework for \\Chord Generation}

\author{Qiqi He}
\affiliation{%
  \institution{NetEase Cloud Music}
  \city{Shanghai}
  \country{China}
}
\email{heqiqi02@corp.netease.com}

\author{Dichucheng Li}
\affiliation{%
  \institution{Individual Researcher}
  \city{Hong Kong}
  \country{China}
}
\email{21210240219@m.fudan.edu.cn}

\author{Xiaoheng Sun}
\affiliation{%
  \institution{NetEase Cloud Music}
  \city{Shanghai}
  \country{China}
}
\email{sunxiaoheng@corp.netease.com}

\author{Anqi Huang}
\affiliation{%
  \institution{NetEase Cloud Music}
  \city{Shanghai}
  \country{China}
}
\email{huanganqi01@corp.netease.com}

\begin{abstract}
Chord generation is an inherently constrained creative task that requires balancing stylistic diversity with music-theoretic feasibility. Existing approaches typically entangle candidate generation and constraint enforcement within a single model, making the diversity–feasibility trade-off difficult to control and interpret.

In this work, we approach chord generation from a system-level perspective, introducing a Retrieval–Edit–Rerank (RER) framework that decomposes the task into three explicit stages: \textit{i) retrieval}, which defines a stylistically plausible candidate space; \textit{ii) editing}, which enforces music-theoretic feasibility through minimal modifications; and \textit{iii) reranking}, which resolves soft preferences among feasible candidates.
This separation provides a controllable pipeline, where each component addresses a distinct aspect of the generation process, thereby enhancing both the interpretability and adjustability of the output chords.

Through objective metrics and subjective evaluation, our decomposed system outperforms all end-to-end chord generation baselines in balancing chord diversity and music-theoretic feasibility. Ablation studies further confirm the complementary roles of each stage in creative exploration and constraint satisfaction.

% Our findings suggest that explicit separation of concerns provides a principled path toward controllable and interpretable symbolic music generation, highlighting system-level design as a viable alternative to single-objective optimization.
\end{abstract}




\maketitle

\section{Introduction}
\label{sec:intro}

Chords, defined as harmonic sets of pitches consisting of multiple notes, play a central role in structuring harmony and enriching musical expression \cite{1969intro}.
% 定义 chord progression generation
Unlike contemporary end-to-end audio generation models like Suno \cite{suno},  the chord generation task operates in the symbolic domain to generate chord sequences. This approach prioritizes editability and DAW interoperability, supporting music production workflows such as melody harmonization for music producers, demanding efficient interaction. % for music creation and post-processing or correction in automatic chord recognition systems. Additionally, chord generation in these application scenarios %

% 难点
% 对比MIR任务
%Unlike conventional music retrieval tasks with clearly defined objectives and accuracy-based evaluation, chord generation is inherently a constrained creative task. It requires not only generating diverse and expressive chord progressions, but also ensuring that the resulting harmonization satisfies music-theoretic constraints with respect to the given melody.
%  Related Work
% 和弦生成任务的大致流派
While Large Language Models (LLMs) and Diffusion Models have dominated music generation \cite{melodyt5, 2026_heartmula}, their adoption for rules-intensive tasks like chord generation remains limited \cite{2024overview}, where probabilistic and deep learning methods still prevail.

% 深度学习之前的传统方法
Before the deep learning era, chord generation heavily relied on probabilistic models and rule-based constraints. Hidden Markov Models (HMMs) were the most representative in this era, evolving from basic trie-structured approaches \cite{2006HMM} to data-driven methods incorporating explicit music theory rules \cite{2007_hmmdb} and interactive user systems \cite{2008microsoft}. 
Beyond HMMs, researchers explored various statistical paradigms to enhance harmonic diversity such as genetic algorithms \cite{2008gene, 2016gene} and Probabilistic Context-Free Grammars (PCFG) \cite{2018_PCFG}.

% 深度学习方法
With the rise of deep learning, neural network-based methods have become the dominant approach for chord generation.
Bi-LSTM-based models were shown to outperform traditional HMM-based methods by better capturing long-range dependencies in musical sequences \cite{2017_blstm}. Subsequent works refined this direction by modeling the conditional dependencies between melody and harmony \cite{2019_clstm}, employing masked sequence learning for inpainting tasks \cite{2021_maskedLSTM}.
Recently, Transformer-based models have been introduced to chord progression generation, modeling global harmonic context \cite{2023_transformer_chord}.  % More recently, more research is dedicated to building Large Language Models for music understanding and generation rather than specialized models for chord generation \cite{melodyt5}.

% 两种方法存在的不足
Despite their differences, these two paradigms exhibit complementary strengths and limitations.
While symbolic and probabilistic models rely on explicit structural representations to ensure harmonic validity, their expressive power and generative diversity are inherently limited.
In contrast, deep learning approaches are capable of capturing stylistic diversity  yet they often require additional mechanisms to satisfy music-theoretic constraints. Existing approaches intertwine candidate generation and constraint enforcement within a single modeling pipeline, making it difficult to simultaneously achieve creative chord progressions and reliable adherence to music-theoretic constraints. 

To address this issue, inspired by the Retrieval-Edit-Rerank (RER) paradigm \cite{2020RERtext}, we adopt a decomposed system design for chord generation. %
    \begin{figure*}[t]
    \centering
    \includegraphics[width=1\textwidth, trim={0.8cm 0.6cm .5cm .6cm}, clip]{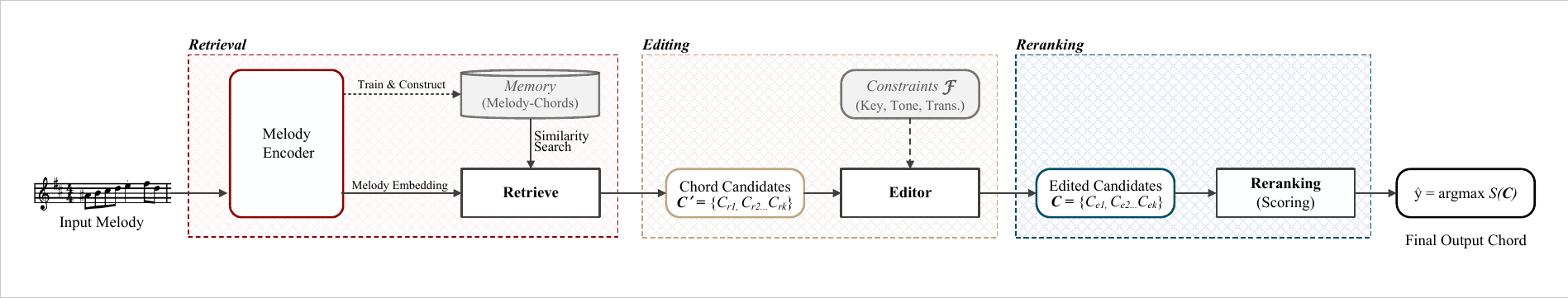}
    \caption{Overview of the proposed Retrieval-Edit-Rerank framework for controllable chord generation.} %Given a melody, the system first retrieves similar chord progressions from the database. Then, a constraint-based editor refines these candidates using music theory. Finally, a reranker scores the candidates to output the optimal chord progression.}
    \label{fig:pipeline}
    \end{figure*} 
% Rather than proposing a new generative model, this work focuses on system-level decomposition and empirical analysis for controllable chord generation. Our contribution lies not in individual components, but in analyzing how this decomposition affects controllability and feasibility in symbolic chord generation. 
Specifically, the proposed system consists of three stages: \textit{i) retrieval}, which defines a stylistically plausible candidate space; \textit{ii) editing}, which enforces music-theoretic feasibility through minimal modifications; and \textit{iii) reranking}, which resolves soft preferences among feasible candidates.  By separating the modeling of expressive diversity from the enforcement of harmonic constraints, our approach aims to achieve both creative generation and theoretical validity in a more stable and controllable manner. %Controllability in the proposed framework arises from this explicit system-level decomposition.

The contributions of this work are threefold:
\begin{enumerate}
  \item We study chord generation under a retrieval--edit--rerank decomposition, explicitly separating candidate selection from symbolic feasibility enforcement.
  \item By applying the RER framework to chord generation, we outperform Transformer-based models, while inherently remaining lightweight in deployment.
  \item We provide empirical analysis and ablation studies demonstrating how this decomposition affects the trade-off between diversity and feasibility.
\end{enumerate}

\section{Method}
%We propose a retrieval–edit–rerank framework for controllable chord generation, which decomposes candidate selection, symbolic feasibility enforcement, and preference resolution into three sequential stages.
As shown in Fig. \ref{fig:pipeline}, given an input melody, the retrieval stage selects a set of stylistically plausible chord progressions from a melody–chord memory without enforcing music-theoretic constraints.
Each retrieved candidate is then processed independently by the editing stage, which enforces harmonic feasibility by projecting it into a constraint-defined feasible space.
Finally, the reranking stage resolves soft preferences among all feasible candidates to select a final chord progression.

The following sections describe each stage in detail.

\subsection{Retrieval}
The retrieval stage aims to identify a set of plausible chord progressions for a given input melody by searching a melody–chord memory, rather than directly generating new sequences.
    \begin{figure}[t]
    \centering
    \includegraphics[width=0.45\textwidth, trim={.65cm 0.65cm .65cm 0.7cm}, clip]{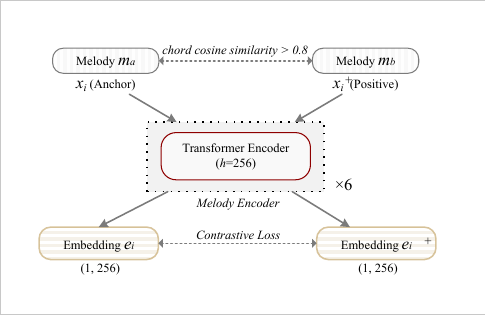}
    \caption{Pipeline of training Melody Encoder.}
    \label{fig:melodyencoder}
    \end{figure} 
    
\textbf{Memory Construction.}
We construct the melody–chord memory through a contrastive learning framework \cite{2020_simclr} as shown in Fig. \ref{fig:melodyencoder}. First, each training melody is clipped into segments according to musical structure. Then, the melody encoder is trained to map these melody segments into a shared embedding space. During training, we form positive pairs $M=\{ m_a , m_b \} $ from melody segments that share the similar chord progression, and negative pairs from segments associated with dissimilar progressions. The encoder is trained with a contrastive loss in \cite{2021_simcse}. After training, every melody clip in dataset is encoded by the trained encoder into an embedding vector of shape $(1, 256)$, and stored in the memory with corresponding chord progression.

\textbf{Retrieval.}
At inference time, the embedding of a query melody is used to retrieve Top $K = 100$ nearest melody–chord candidates using cosine similarity. To ensure scalability, we employ FAISS \cite{2019_faiss} for embedding retrieval. The retrieved candidates are passed to the editing stage.

By restricting retrieval to stylistic plausibility rather than correctness, this stage provides diverse yet structurally meaningful starting points for controllable chord generation.
%The training process of encoder encourages melody–chord pairs with similar harmonic contexts to be close in the embedding space, while pushing apart dissimilar pairs. Here, harmonic context is treated as a weak supervisory signal rather than an explicitly modeled music-theoretic concept. This design allows the retrieval stage to focus exclusively on modeling melodic similarity and stylistic correspondence, leaving the enforcement of harmonic feasibility to subsequent stages.

\subsection{Editing}
\label{sec:edit}

While the retrieval stage captures high-level stylistic diversity, raw retrieved chord progressions may violate fundamental musical theory due to the uninterpretability of the representation. To address this, we formulate the editing stage as a projection optimization problem. The goal is to project the retrieved candidate $C_{r}$ onto a feasible set $\mathcal{F}$ that satisfies the basic constraints of music theory.

Formally, let $\mathcal{V}$ denote the vocabulary of chords. The feasible space $\mathcal{F} \subset \mathcal{V}$ is defined by a set of music-theoretic constraints derived from musical theory, ensuring the generated sequence possesses basic structural integrity. These priors include:

\begin{enumerate}
\item \textit{Tonal Alignment (Vertical):} Ensures harmonic compatibility between the generated chords and the input melody.
\item \textit{Cadential Resolution (Horizontal):} Enforces proper initiation and cadential constraints of chord progressions, particularly at phrase boundaries.
\item \textit{Global Regularization:}  Filters out harmonically implausible chords while permitting stylistically intentional dissonance (e.g., in jazz) according to previous research \cite{2008gene}.
\end{enumerate}

 The editing process finds a feasible candidate $C_e \in \mathcal{F}$ that best preserves the style of the retrieved candidate $C_r$ by minimizing a modification cost:

\begin{equation}
\label{eq:editor}
    C_{e} = \operatorname*{argmin}_{C \in \mathcal{F}} \; d(C, C_{r})
\end{equation}

% where the harmonic distance metric $d(\cdot, \cdot)$ and detailed constraint specifications and parameter settings in our implementation are provided\footnote{https://github.com/RERChordGen/Editor}.
Here, the distance $d(\cdot, \cdot)$ is formulated as a weighted sum of penalties corresponding to the three aforementioned constraints in $\mathcal{F}$ (e.g., rewarding authentic cadences like $V \to I$ at phrase boundaries), with a style parameter to tolerate genre-specific complexity. Since the constraints in $\mathcal{F}$ can be modeled as transition and emission probabilities, we define $\mathcal{V}$ as a set of 48 categories as previous works \cite{2023_transformer_chord} and solve Eq. \ref{eq:editor} efficiently using the Viterbi algorithm. 

Fig. \ref{fig:editor}b illustrates the process of projecting the retrieved candidates $C_{r}$ in Fig. \ref{fig:editor}a onto the feasible set $\mathcal{F}$ during the editing stage. Editing stage provides a constraint projection allowing the retrieval stage to explore potentially risky stylistic chords, while the projection guarantees that the final output remains within musical plausibility.

\subsection{Reranking}
\label{sec:rerank}
    \begin{figure}[t]
    \centering
    \includegraphics[width=0.48\textwidth, trim={.65cm 0.65cm .65cm 0.6cm}, clip]{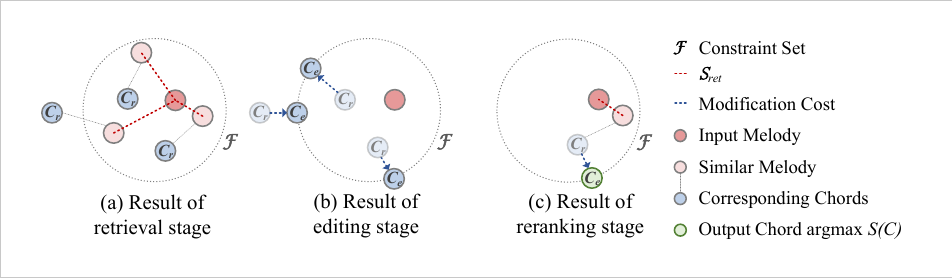}
    
    \caption{Results of each stage. Each retrieved candidate $C_r$ is projected onto a music‑theoretic constraint set $\mathcal{F}$, resulting in a valid progression $C_e$ in (b). Then in (c), the reranking stage selects the final output by minimizing $S(C)$ defined over a retrieval and editing subscores, as detailed in Sec.~\ref{sec:rerank}.}
    \label{fig:editor}
    \end{figure} 
After music-theoretic feasibility is enforced by the editing stage, a set of feasible chord progressions exist for a given melody.
The reranking stage aims to select a final solution by resolving soft preferences among these feasible candidates, without introducing additional hard constraints.

Each feasible candidate $C$ is assigned a global ranking score $S(C)$ to explicitly balance the stylistic guidance from the retrieval stage and the theoretical compliance from the editing stage:
\begin{align}
    S(C) = \lambda \cdot \mathcal{S}_{\text{ret}}(m, m') + (1 - \lambda) \cdot \mathcal{S}_{\text{edit}}(d)
    \label{eq:global_score}
\end{align}
where $\lambda \in [0, 1]$ is a hyperparameter controlling the relative importance of the two components, which are defined as follows:

\begin{enumerate}
    \item \textit{Retrieval Score} $\mathcal{S}_{\text{ret}}$ measures the similarity between the query melody $m$ and the retrieved melody $m'$, indicating stylistic relevance inherited from the retrieval stage;

    \item \textit{Editing Score} $\mathcal{S}_{\text{edit}}$
    indicates the editing cost $d$ in Sec. \ref{sec:edit}. A lower cost yields a higher score, promoting fidelity to the raw retrieval:
    \begin{align}
        \mathcal{S}_{\text{edit}}(d) = \frac{2}{1 + \exp(\gamma \cdot d)}
        \label{eq:sigmoid_decay}
    \end{align}
    where $\gamma$ is a scaling factor. 
\end{enumerate}

 These two components are shown in Fig. \ref{fig:editor}c. Moreover, $\lambda$ is calculated across all experiments, determined via a grid search on the validation set to best balance stylistic relevance with harmonic feasibility.% ranking feasible candidates according to .
%The modification cost, in particular, 

The final chord progression is selected by $\arg \max S (C)$ from feasible candidates. This score %is used to express a preference for preserving the structure of the retrieved candidate while the editing stage guarantees feasibility. It 
biases selection toward solutions with lower modification cost without overriding diversity.
%, but it is not enforced as a requirement.
% It serves not as the primary objective (which is handled by editing) but as a tie-breaking criterion to further promote structural fidelity among %otherwise equally feasible candidates.
By separating feasibility enforcement from soft preference resolution, the reranking stage complements the retrieval and editing stages while maintaining a clear system-level decomposition. This separation significantly enhances the interpretability of the generated chord progressions.% Importantly, all components in reranking stage are treated as soft preferences rather than constraints.

\noindent In practice, controllability can be achieved by adjusting system-level interfaces such as the size of the retrieved candidate set, the constraint set $\mathcal{F}$, or the preference weights in reranking.
These mechanisms allow the system to adapt to different harmonic conventions and diversity–feasibility trade-offs.
\section{Experiments}

\subsection{Datasets \& Metrics}
\textbf{Datasets.}
We use four datasets: \texttt{RWC-Pop} \cite{2002rwc}, a paid pop-song dataset with AIST \cite{2006aist} annotations; \texttt{Wikitest}, derived from the test split of Wikifonia \cite{2017_blstm}; \texttt{POP-909} \cite{2020pop909}, containing 909 professionally created pop songs; and \texttt{Nk1k3}, an internal dataset with 1,558 labeled pop songs.

For retrieval-stage training, we use 2,465 songs from \texttt{POP-909} and \texttt{Nk1k3}, split 80\%/20\% for training and validation, and clip each song into 16-bar segments.

\noindent \textbf{Metrics.}
For evaluation, we use 543 songs from \texttt{RWC-Pop} and \texttt{Wikitest}, yielding 4,834 melody clips with no overlap with the training data.

For objective evaluation, we adopt metrics from prior work \cite{2021metric}: CHE, CC, and CTD for diversity and transition structure, and PCS, MCTD, and CTnCTR for harmonic compatibility and tonal fitness.

For subjective evaluation, following \cite{2022_accomp}, we use a five-point scale on Harmonicity, Creativity, and Overall preference. We randomly select 15 test clips (8--16 bars) and recruit 30 participants, including 7 professionals, 15 musically trained amateurs, and 8 non-musicians.

\subsection{Ablation Study}
\begin{table}[]
\centering
\scriptsize
\scalebox{0.95}{
\begin{tabular}{l|l|cccccc}
\toprule[1pt]
Dataset & Method 
& $\Delta$CHE $\downarrow$
& $\Delta$CC $\downarrow$
& $\Delta$CTD $\downarrow$
& PCS $\uparrow$ 
& MCTD $\downarrow$ 
& \makecell[c]{CTn\\CTR$\uparrow$} \\
\hline\hline

\multirow{6}{*}{\texttt{RWC-Pop}}
 & Ground Truth  & 1.4131 & 32.5385 & 0.8532 & 0.9978 & .1318 & .4216 \\
\cline{2-8}
 & \textit{RER Frame}       
 & \textbf{+.0579} & \textbf{-0.2418} & \textbf{-.2083}
 & \textbf{1.4398} & .1280 & \textbf{.4781} \\
 & \textit{W/o Retrieval} 
 & -.5912 & -13.0000 & -.5150
 & 1.3888 & \textbf{.1259} & .4751 \\
 & \textit{W/o Editor}    
 & -.4035 & -10.0330 & -.2454
 & 1.2917 & .1307 & .4671 \\
 & \textit{W/o Reranking} 
 & -.2025 & -6.4506 & -.2358
 & 1.3633 & .1292 & .4640 \\
 & \textit{Random}        
 & +.6959 & +28.6044 & +.2737
 & 0.6288 & .1367 & .3454 \\
\hline
\multirow{6}{*}{\texttt{Wikitest}}
 & Ground Truth  & 1.2912 & 19.3926 & 0.4536 & 0.9307 & .0887 & .4206 \\
\cline{2-8}
 & \textit{RER Frame}       
 & \textbf{-.0358} & \textbf{-0.5989} & \textbf{+.1774}
 & \textbf{0.7601} & \textbf{.0864} & .3839 \\
 & \textit{W/o Retrieval} 
 & -.9402 & -8.5932 & -.3854
 & 0.7554 & .0877 & \textbf{.3856} \\
 & \textit{W/o Editor}    
 & -.6922 & -11.4041 & +.5830
 & 0.7489 & .0887 & .3749 \\
 & \textit{W/o Reranking} 
 & -.0385 & -2.0430 & -.2846
 & 0.7496 & .0891 & .3772 \\
 & \textit{Random}        
 & +.8813 & +26.4985 & +.5053
 & 0.7122 & .0906 & .3657 \\
\hline
\end{tabular}
}
\caption{Ablation results on \texttt{RWC-Pop} and \texttt{Wikitest}. Best results are highlighted. "W/o" denotes removal of a specific stage.}
\label{tab:as}
\end{table}

We conduct an ablation study by removing each stage of our framework independently (Table \ref{tab:as}). Removing retrieval (operating on unconstrained candidates) leads to a substantial collapse in chord diversity and coverage (CHE, CC), confirming that retrieval is crucial for providing a stylistically diverse candidate space. The harmonic transitions also become overly conservative (lower CTD).
Conversely, removing the editor results in a clear degradation in melody-chord compatibility (PCS, MCTD, CTnCTR), verifying that constraint projection is essential for enforcing feasibility of chord progressions. Moreover, without editing, harmonic transitions become unstable and excessive (higher CTD).
Removing reranking has a minimal impact on feasibility metrics but leads to a slight, consistent drop across several preference-related scores. This supports its designated role as a soft preference resolver that refines the selection among already-feasible candidates.%, rather than a core feasibility mechanism.

Critically, the opposing effects on CTD highlight the complementary roles of the retrieval and editing stages. Removing retrieval makes progressions overly conservative, whereas removing editing makes them overly erratic. All variants significantly outperform the random baseline, confirming that the evaluation metrics capture non-trivial harmonic structure.
    \begin{figure}[h]
    \centering
    \includegraphics[width=0.48\textwidth, trim={0.65cm 0.65cm .65cm 0.7cm}, clip]{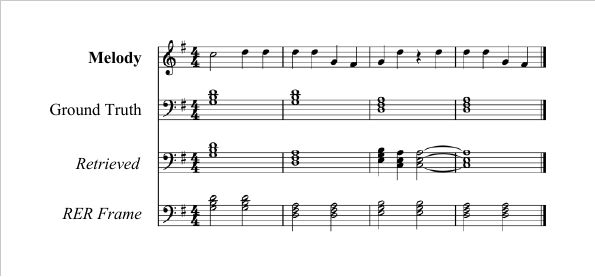}
    \caption{A bad case study where the \textit{Retrieved} candidate is distant from the input melody. The proposed \textit{RER Frame} results in a more conservative chord progression to ensure harmonic stability.}
    \label{fig:sheet}
    \end{figure} % Badcase study
    
We also analyze failure cases to assess system robustness. As shown in Fig. \ref{fig:sheet}, the system occasionally (2\textperthousand \space in evaluation) retrieves chord sequences that are distant from the constraint set $\mathcal{F}$. In such cases, the editing stage applies substantial modifications to satisfy feasibility constraints, causing the final output to rely more heavily on the editing stage. Consequently, the progression tends to become conservative to prioritize harmonic correctness.

\subsection{Comparison with Prior Works}
\begin{table}[]
\centering
\scriptsize
\scalebox{0.9}{
\begin{tabular}{l|l|cccccc}
\toprule[1pt]
Dataset & Method 
& $\Delta$CHE $\downarrow$& $\Delta$CC $\downarrow$& $\Delta$CTD 
$\downarrow$& PCS  $\uparrow$ & MCTD $\downarrow$ & \makecell[c]{CTn\\CTR$\uparrow$} \\
\hline\hline
\multirow{4}{*}{\texttt{RWC-Pop}}
 & Ground Truth  
 & 1.4179 & 24.6207 & .8590 
 & 0.7896 & .1322 & .4349 \\
\cline{2-8}
 & \textit{RER Frame} 
 & \textbf{-.2048} & \textbf{-4.9081} & \textbf{-.2160}
 & \textbf{1.0764} & .1262 & \textbf{.4824} \\
 & \textit{TransformerLM} \cite{2023_transformer_chord} 
 & -.4440 & -7.3678 & -.3294
 & 1.0192 & \textbf{.1285} & .4181 \\
 & \textit{Bi-LSTM}\cite{2017_blstm}   
 & -.3624 & -8.7619 & -.4310
 & 0.9465 & .1469 & .4814 \\
 & \textit{HMM}\cite{2006HMM}      
 & -.5041 & -8.3563 & -.2475
 & 0.6527 & \textbf{.1285} & .4732 \\
\hline
\multirow{4}{*}{\texttt{Wikitest}}
 & Ground Truth  
 & 1.2919 & 19.2980 & .4540 
 & 0.9218 & .0884 & .4195 \\
\cline{2-8}
 & \textit{RER Frame} 
 & \textbf{-.0372} & \textbf{-0.6189} & +.1777
 & 0.9432 & \textbf{.0884} & \textbf{.4835} \\
 & \textit{TransformerLM} \cite{2023_transformer_chord} 
 & -.1423 & -1.4957 & \textbf{-.0257}
 & \textbf{1.3067} & .0929 & .4622 \\
 & \textit{Bi-LSTM}\cite{2017_blstm}   
 & -.1178 & -2.4072 & -.0484
 & 0.5511 & .0977 & .4624 \\
 & \textit{HMM}\cite{2006HMM}       
 & -.3163 & -2.6963 & +.3784
 & 0.7226 & .0889 & .3777 \\
\hline
\end{tabular}
}
\caption{Results of RER-Frame and baselines on two datasets. Music segments that do not satisfy the baseline generation conditions are excluded, thus the ground truth used in this table differs from the ablation study.}
\label{tab:comp}
\end{table}

\textbf{Objective Evaluation.} Table \ref{tab:comp} reveals imbalanced trade-offs across baselines. \textit{HMM} favors harmonic stability (low MCTD) but produces overly conservative progressions (low CHE/CC). \textit{Bi-LSTM} captures local context but lacks global constraints, resulting in looser transitions (highest MCTD). \textit{TransformerLM} \cite{2023_transformer_chord} performs well on local relationships, with strong CTD and PCS, but lower CHE and CC, suggesting a tendency to overfit frequent progressions rather than balance diversity and feasibility.

In contrast, the proposed RER framework achieves a more balanced result. It restores harmonic diversity (CHE and CC closest to Ground Truth) while maintaining competitive structural stability and local compatibility. This advantage comes from the explicit decomposition of retrieval, editing, and reranking. Moreover, RER outperforms Transformer-based models without additional GPU cost.

    \begin{figure}[t]
    \centering
    \includegraphics[width=0.48\textwidth, trim={.65cm 0.65cm .65cm 0.7cm}, clip]{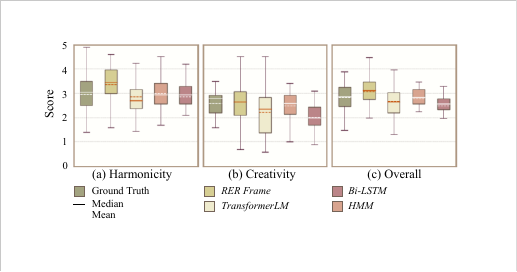}
    \caption{Visualization of boxplots for subjective evaluation. A shorter distance between two caps indicates a more stable performance of the model.}
    \label{fig:comp}
    \end{figure} % Baselines Comp. 

\textbf{Subjective Evaluation.} A listening study (Fig. \ref{fig:comp}) supports the objective findings. While the baselines tended toward either rigid correctness (\textit{HMM}) or randomness (\textit{Bi-LSTM} and \textit{TransformerLM}), RER achieved the most balanced scores in Harmonicity, Creativity, and Overall preference. This suggests that our decomposition strategy produces outputs that are both statistically closer to the Ground Truth and more perceptually coherent.

\section{Conclusion}
We presented a retrieval--edit--rerank framework for controllable chord
generation that explicitly separates stylistic candidate selection,
music-theoretic feasibility enforcement, and preference-based selection.
Through objective and subjective evaluations, we showed that this
decomposition provides a more balanced trade-off between harmonic
diversity, feasibility, and perceptual quality than end-to-end chord
generation models. These findings suggest that explicit system-level
decomposition is a practical alternative to monolithic generation for
constraint-sensitive symbolic music tasks. Future work will investigate a
global controller over the three stages to reduce manual tuning while
preserving the benefits of decomposition.

% \begin{acks}
% \end{acks}


\bibliographystyle{ACM-Reference-Format}
\bibliography{icmr2026}

\end{document}